\documentclass[apj]{emulateapj}

\usepackage{amsmath}
\usepackage{color}
\usepackage{needspace}
\usepackage{gensymb}
\usepackage{enumitem}

\newcommand{\angstrom}{\textup{\AA}}
\newcommand{\Halpha}{{H$\alpha$}}
\newcommand{\Hbeta}{{H$\beta$}}
\newcommand{\CaII}{{\ion{Ca}{2}}}
\newcommand{\SiIV}{{\ion{Si}{4}}}

\newcommand{\Heteneightthirty}{{\ion{He}{1}  $10830$\,$\angstrom$}}

\shorttitle{Surges and \SiIV \ bursts: observations and synthesis}
\shortauthors{N\'obrega-Siverio, D. et al.}

\begin{document}

%
\title{Surges and \SiIV \ bursts in the solar atmosphere. Understanding IRIS and SST observations through RMHD experiments}

%
\author{D. N\'obrega-Siverio \altaffilmark{1,2}, J. Mart\'inez-Sykora\altaffilmark{3,4}, F. Moreno-Insertis\altaffilmark{1,2}, and L. Rouppe van der Voort\altaffilmark{5}}

\affil{$^1$ Instituto de Astrofisica de Canarias, Via Lactea, s/n, E-38205 La Laguna (Tenerife), Spain}
\affil{$^2$ Department of Astrophysics, Universidad de La Laguna, E-38200 La Laguna (Tenerife), Spain}    
\affil{$^3$ Lockheed Martin Solar and Astrophysics Laboratory, Palo Alto, CA 94304, USA}
\affil{$^4$ Bay Area Environmental Research Institute, Petaluma, CA, USA}
\affil{$^5$ Institute of Theoretical Astrophysics, University of Oslo, P.O. Box 1029 Blindern, NO-0315 Oslo, Norway}
\email{dnobrega@iac.es, juanms@lmsal.com, fmi@iac.es, rouppe@astro.uio.no}

%
\begin{abstract}
Surges often appear as a result of the emergence of magnetized plasma from the solar interior. Traditionally, they are observed in chromospheric lines such as \Halpha\ 6563 $\angstrom$ and  \CaII \ 8542 $\angstrom$. However, whether there is a response to the surge appearance and evolution in the \SiIV\ lines or, in fact, in many other transition region lines has not been studied. In this paper we analyze a simultaneous episode of an \Halpha\ surge and a \SiIV\ burst that occurred on 2016 September 03 in active region AR12585. To that end, we use coordinated observations from the Interface Region Imaging Spectrograph (IRIS) and the Swedish 1-m Solar Telescope (SST). For the first time, we report emission of \SiIV\ within the surge, finding profiles that are brighter and broader than the average. Furthermore, the brightest \SiIV\ patches within the domain of the surge are located mainly near its footpoints.  To understand the relation between the surges and the emission in transition region lines like \SiIV, we have carried out 2.5D radiative MHD (RMHD) experiments of magnetic flux emergence episodes using the Bifrost code and including the non-equilibrium ionization of silicon. Through spectral synthesis we explain several features of the observations. We show that the presence of \SiIV\ emission patches within the surge, their location near the surge footpoints and various observed spectral features are a natural consequence of the emergence of magnetized plasma from the interior to the atmosphere and the ensuing reconnection processes. 
\end{abstract}


%
\Needspace{5\baselineskip}
\section{Introduction}\label{sec:1}

Surges are a good example of the complexity of chromospheric ejections. In \Halpha\ 6563 $\angstrom$,  surges are seen as darkenings in the blue/red wings of the line with line-of-sight (LOS) apparent velocities of a few to several tens of km s$^{-1}$ on areas with projected lenghts of $10-50$ Mm (\citealp{Kirshner1971}, \citealp{roy1973}, \citealp{Cao1980}, \citealp{Schmieder1984}, \citealp{Chae1999}, \citealp{Guglielmino:2010lr}, among others). They can be recurrent \citep{Schmieder1995, Gaizauskas,Jiang2007, Uddin2012} and have apparent rotational and helical motions \citep{Gu1994, canfield1996, Jibben2004, Bong:2014}. Recent observations show that the surges consist of small-scale thread-like structures \citep{nelson2013, Li2016} and that appear to be related to shocks \citep{YangH:2014} and Kelvin-Helmholtz instabilities \citep{Zhelyazkov2015}. The surges have also been detected in other chromospheric lines such as the \CaII \ 8542 $\angstrom$ infrared triplet \citep{Yang:2013h, Kim:2015y}, \Heteneightthirty \ \citep{Vargas2014}, \Hbeta\ 4861 $\angstrom$ \citep{Zhang:2000hb,Liu2004}, and also in \CaII\ H \& K \citep{Rust1976,Nishizuka:2008zl,liu2009}.

Surges normally appear related with emerging flux regions \citep[EFR,][]{Kurokawa2007}. This is especially evident in active regions (AR), where there are many observations of such cool ejections \citep[e.g.,][]{Brooks2007, Madjarska2009, Wang2014}. Furthermore, surges are also associated with other phenomena such as light bridge \citep{Asai2001,Shimizu:2009,Robustini:2016} and explosive events (EEs), EUV or X-ray ejections \citep[see,  e.g.,][]{Schmahl1981, canfield1996, Chen2008, Madjarska2009, Zhang2014}. It has also been suggested that surges can be related to Ellerman bombs \citep[EBs;][]{watanabe2011,Yang:2013h,Vissers:2013}; nevertheless, this relation does not seem to be common \citep{Rutten:2013}.

On the other hand, observations carried out with the IRIS satellite \citep{De-Pontieu:2014vn} have provided a new perspective on transient phenomena in the chromosphere and transition region. In particular, the connection between UV bursts observed with IRIS \citep[also refered as IRIS bursts or IRIS bombs;][]{Peter:2014h} and previously known phenomena has become an active area of research \citep[see, e.g., ][among others]{Judge:2015p,Vissers:2015,Tian:2016h,Grubecka:2016}. In this context, recent papers report on the coexistence of surges with bursts in \SiIV \ (\citealp{Kim:2015y}, \citealp{Huang:2017z} and \citealp{Madjarska:2017}); nevertheless, the focus of those papers was mainly on the bursts, without presenting an in-depth analysis of the associated surges.

\begin{figure*}
\epsscale{1.18}
\plotone{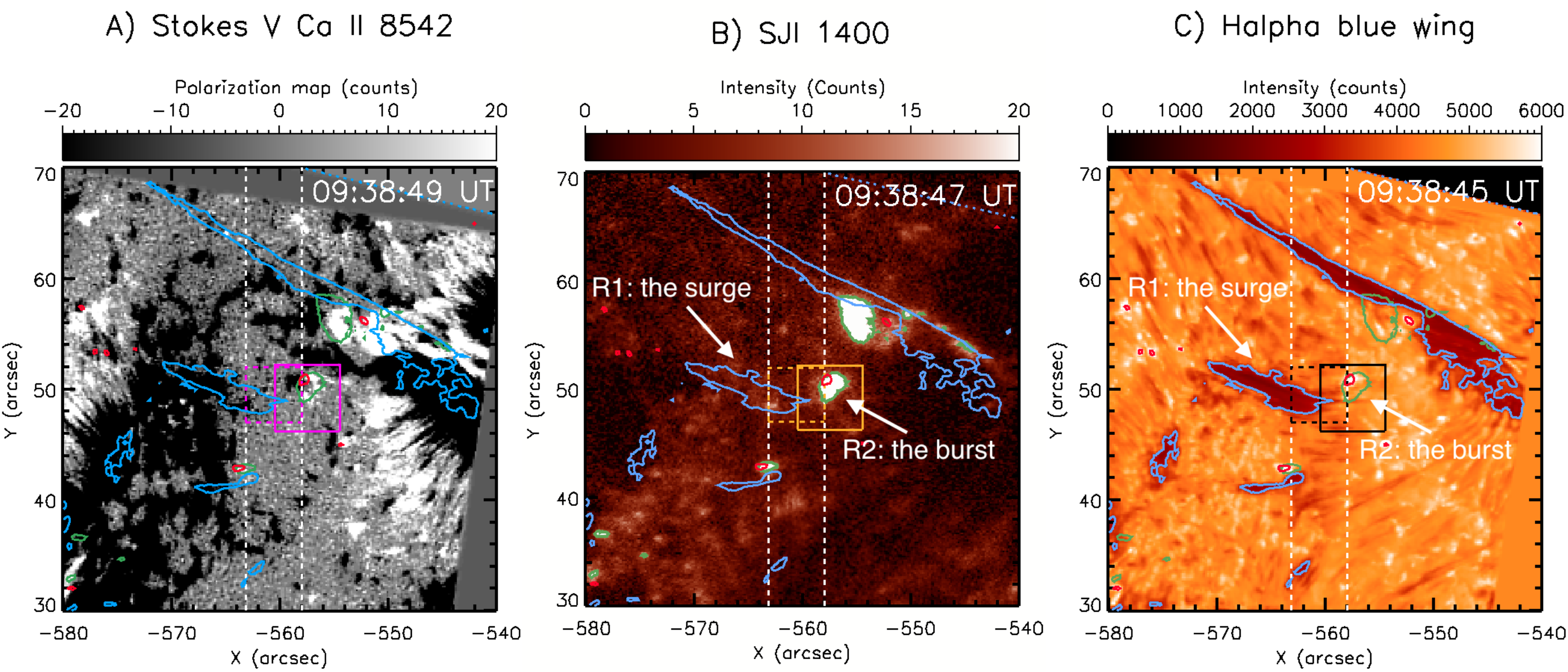}
\caption{Context images of the active region AR12585 on 2016 September 03 at $\sim$ 09:39 UT observed by IRIS and SST. (A) Stokes V polarization map of the far wings of \CaII \ 8542 $\angstrom$ serving as a proxy of the photospheric magnetic field. (B) IRIS slit-jaw image (SJI) in the \SiIV \ 1400 $\angstrom$ passband. (C) SST/CRISP image in the blue wing of the \Halpha \, at -46~km~s$^{-1}$. Bright regions in the SJI 1400 ($> 16$ counts) are overplotted as green contours in all the three panels, while the bright ($6500$ counts) and dark structures ($3100$ counts) of the \Halpha \ blue wing scan are superimposed in red and blue respectively to facilitate the identification. The arrows R1 and R2 point to the different regions, surge and burst respectively, explained in detail in Section \ref{sec:3}. The region used for the integration for Figure \ref{obsfig2} is marked out with a solid rectangle. White dashed lines indicate the region covered by the IRIS raster. The area contained in the dashed rectangle is used for Figures \ref{obsfig3} and \ref{obsfig4}. \\
(An animation of the figure is available showing the time evolution of the three panels from $\sim$ 08:56 UT to the end of the coordinated observations at $\sim$ 10:03 UT.) \label{obsfig1}}
\end{figure*}

The aim of this paper is to explore the response of transition region lines during surge formation and characterize and explain the corresponding spectral profiles. Furthermore, we want to provide theoretical support for the relation between \Halpha \ surges and other phenomena such as \SiIV \ bursts and EBs. To that end, we use coordinated observations from space and the ground with IRIS and SST \citep{Scharmer:2003ve}. Those instruments provide high-resolution observations that cover chromospheric and transition region lines, which are essential for this study. Theoretical support is given by 2.5D numerical experiments similar to the one discussed in detail by \cite{Nobrega-Siverio:2016}, hereafter NS2016, but extending the Bifrost code capabilities \citep{Gudiksen:2011qy}  by including a module developed by \cite{olluri:2013aa} that calculates the ionization state of the \SiIV\ in non-equilibrium (NEQ).

The layout of the paper is as follows. Section \ref{sec:2} describes the IRIS and SST coordinated observations, and briefly the numerical models. In Section \ref{sec:3}, we analyze the observations showing mainly \SiIV \ profiles, Doppler shifts and line widths in the region of interest. In Section \ref{sec:4} we study the synthetic spectral profiles obtained from the numerical experiments comparing them with the observations. Finally, Section \ref{sec:5} contains the conclusions and summary.

%
\Needspace{5\baselineskip}
\section{Observations and numerical model}\label{sec:2}

%
%
\Needspace{5\baselineskip}
\subsection{Coordinated observations}\label{sec:2.1}

For this paper we have used coordinated observations of IRIS
and SST obtained on 2016 September 03 from 07:49:21 to 10:03:37 UT centered on active region AR12585 at heliocentric coordinates $(x,y) = (-561\arcsec,44\arcsec)$ and observing angle $\mu=0.8$.

\begin{itemize}

\item The IRIS dataset corresponds to the observation program OBSID: 3625503135. Those observations cover a maximum field of view (FOV) of $65 \arcsec \times 60 \arcsec$. The spectral information is obtained through a medium-dense raster that covers $60\arcsec$ in the $y$ direction and scans $5\farcs28$ in the $x$ direction in 16 steps of $0\farcs33$ separation for each step, with a raster cadence of $20$ s. The exposure time is $0.5$ s, the spatial scale is 0\farcs166 pixel$^{-1}$, and the spectral sampling of the FUV spectra has been binned four times in the camera from 2.7~km~s$^{-1}$ to 11~km~s$^{-1}$ to increase the signal level. Concerning the slit-jaw images (SJI), \ion{C}{2} 1330, \SiIV \ 1400 and \ion{Mg}{2} k 2796 $\angstrom$ were taken with a cadence of $10$ s obtaining a total of 800 images in each passband. The data have been calibrated for dark current, flat-field and geometric correction. We analyse level 2 data \citep {De-Pontieu:2014vn} produced by SolarSoft's $\texttt{iris\_prep.pro}$ routine version 1.83 which includes an updated version of the wavelength calibration (level2 version L12-2017-04-23). When studying IRIS FUV spectra profiles in the following, we shall consider temporal and spatial evolution of features like intensities, relative velocities, line widths, and asymmetries. On the other hand, the absolute wavelength calibration of these observations is not good enough to allow for a determination with accuracy better than 20-25 km s$^{-1}$ of the global Doppler shift. This is due to the lack of photospheric lines in this data set as a result of the short exposures.

\item The SST dataset contains \CaII \ 8542 $\angstrom$ and \Halpha \ 6563 $\angstrom$ spectral scans obtained with the SST/CRISP Fabry-P\'erot Interferometer \citep{Scharmer:2006}. The CRISP scans have a temporal cadence of $20$ s, a spatial sampling of 0\farcs057 per pixel and a FOV of $\sim 58\arcsec \times 58\arcsec$. The diffraction limit of the SST at 6563~\AA\ is 0\farcs14. H$\alpha$ was sampled at 15 line positions from $-1500$~m\AA\ to $+1500$~m\AA\ (FWHM transmission profile: 66 m\AA), and \CaII\ 8542  was sampled in spectropolarimetric mode at 21 spectral positions between $-1750$ m$\angstrom$ and $+1750$ m$\angstrom$ offset from the line core. The wings were sparsely sampled and the line core region was sampled with equidistant steps of $70$ m$\angstrom$ between $\pm$ $455$ m$\angstrom$ (FWHM transmission profile: $107$ m$\angstrom$). Taking into account the magnetic sensitivity and the wide range in formation height of the \CaII\ 8542  line (see, e.g., \citealp{delaCruz:2010}), we construct photospheric polarization maps by adding the Stokes V maps of the 3 outer spectral positions in both wings, preserving the sign of magnetic polarity by effectively subtracting the sum of the 3 red wing maps from the sum of the 3 blue wing maps. While the \CaII \ 8542 line core originates from the chromosphere, the far wings originate from the photosphere, so these accumulated polarization maps serve as a proxy for the photospheric magnetic field (\citealp{delaCruz:2012,ortiz2014}). The CRISP data were processed following the CRISPRED data reduction pipeline \citep{delaCruz:2015}, which includes Multi-Object Multi-Frame Blind Deconvolution image restoration \citep{van-Noort:2005uq}. The alignment with the IRIS data was done by scaling down the CRISP data to the IRIS pixel-scale and through cross-correlation of the \CaII \ 8542 $\angstrom$ wing images with the IRIS SJI 2796 channel. This alignment has been shown to be accurate down to the level of the IRIS pixel (0\farcs17, \citealp{Rouppe:2015}).

\end{itemize}

Figure \ref{obsfig1} shows a fraction of the FOV of the region observed by IRIS and SST at $\sim$09:39 UT. Panel A shows a polarization map of the \CaII \ 8542 $\angstrom$ wings that serves as a proxy of the photospheric magnetic field in active region AR12585. Panel B contains the IRIS SJI 1400; and panel C is the SST image in the blue wing of \Halpha \ at $-46$~km~s$^{-1}$ (an accompanying movie of the figure is provided in the online version). The solid rectangle in all three panels identifies the area of interest where the events originate. Additionally, part of that region is also covered by the IRIS raster (see area within the vertical dashed lines in the figure), so we have detailed \SiIV \ spectral information from part of the surge and burst to compare with the synthetic spectra of the numerical experiment. In panel B, the arrows mark two regions, R1 and R2, that correspond to a surge and a burst, respectively, and which are explained in detail in Section \ref{sec:3}. At the time of the figure, approximately at 09:39 UT, two opposite polarities are colliding within the solid rectangle (see panel A) while in panel B a \SiIV \ burst (green contour) is visible around $(x,y) = (-557\arcsec,51\arcsec)$. Co-spatial with the upper part of the bright region in the SJI 1400 at $(x,y) = (-558\arcsec,51\arcsec)$, we identify an intense brightening in the \Halpha\ wing in panel C, which is highlighted with a red contour within the burst. The \Halpha\ spectral profile reveals that this is an EB (see Section \ref{sec:3}). The blue contours in the three panels mark out dark structures visible in panel C. Some of those dark structures correspond to \Halpha \ surges; in particular, the relevant one for this paper is marked by the R1 arrow and appears during the \SiIV \ burst (R2). Following the darkenings in the SST images, it seems that this surge is ejected leftwards of $(x,y) \sim (-559\arcsec,49\arcsec)$; we will refer to that location as \textit{the footpoint of the surge} in the following.

%
%
\Needspace{5\baselineskip}
\subsection{The numerical experiment}\label{sec:2.2}

In order to understand the physics of the surges and related phenomena, it is necessary to provide theoretical support to the observations. To that end, we have run two 2.5D numerical experiments in which a surge results from magnetic flux emergence and where thermal conduction and radiative transfer are treated in a self-consistent manner.

\begin{itemize}

\item Concerning the code, the present experiments have been performed with the 3D RMHD Bifrost code \citep{Gudiksen:2011qy, Carlsson:2012uq, Hayek:2010ac}. In addition, we have enabled a module to calculate the non-equilibrium (NEQ) ionization for silicon developed by \cite{olluri:2013aa}. Thus, we are able to compute the emissivity and then the synthetic spectra under optically thin conditions, as in the papers by \cite{olluri:2015} and \cite{Martinez-Sykora:2016obs}, for different orientations of the LOS ($-15$, $0$, and $15$ degrees) with respect to the vertical direction $z$.

\item Regarding the numerical setup, the two runs are similar to the one in the NS2016 paper. The main differences are the following: 1) we have kept the resolution of NS2016 but increased the physical domain, namely, $0.0$~Mm $\leq x \leq$ $32.0$~Mm and $-2.6$~Mm $\leq z \leq$ $30.0$~Mm, where $z=0$~Mm corresponds to the solar surface; 2) one of the current experiments has a vertical coronal magnetic field while the other is slanted; and 3) the initial axial magnetic flux is lower than in NS2016, $\Phi_0 = 6.3 \times 10^{18}$~Mx. More results about the numerical experiment will be provided in a follow-up paper \citep{Nobrega-Siverio:2017b}, where a deeper analysis of the theoretical aspects of the experiments is carried out.

\item Since we are interested in synthesizing \SiIV \ for comparison purposes with the IRIS observations, spatial and spectral PSF (gaussian) degradation of the theoretical data must be carried out to reduce the spectral profiles into the IRIS instrumental spatial and spectral resolution ($0.33''$ and $26$ m\angstrom \  respectively). Finally, we convert the degraded line intensity  from CGS units (erg s$^{-1}$ cm$^{-2}$ sr$^{-1}$) to IRIS photon count number (DN s$^{-1}$) for direct comparisons using
\begin{eqnarray}
	I_{_{{IRIS}}} & = &  I_{_{{CGS}}}\, \frac{A\, p\, w}{r^2 } \frac{\lambda}{h\, c} \frac{1}{k}
\label{eq:dn}
\end{eqnarray}
where $A = 2.2$ cm$^2$ is the effective area for wavelengths between $1389$ and $1407$ \angstrom, $p=0\farcs167$ is the spatial pixel size, $w=0\farcs33$ is the slit width, $r=3600 \cdot 180/\pi$ is the conversion of arcsec to radians, $\lambda=1402.77$ \angstrom \ is the wavelength of interest, $h$ is the Planck constant, $c$ the speed of light, and $k=4$ is the number of photons per DN in the case of FUV spectra \citep[see][for the instrumental specifications about IRIS]{De-Pontieu:2014vn}.

\end{itemize}

%
\Needspace{5\baselineskip}
\section{Observations: Results}\label{sec:3}

Along the observation, we are able to distinguish two main episodes of surge activity in the SST/CRISP maps with accompanying brightenings in the SJI 1400: one taking place between 08:14 and 08:39 UT and the other between 09:27 UT and 10:03 UT, which is the end of the coordinated observation. Both episodes seem to be related to the appearance of positive polarity patches within a preexisting negative polarity region. In this paper we focus on the second one since the IRIS spectral raster covers most of it in contrast to the first episode.

\begin{figure}
\epsscale{1.15}
\plotone{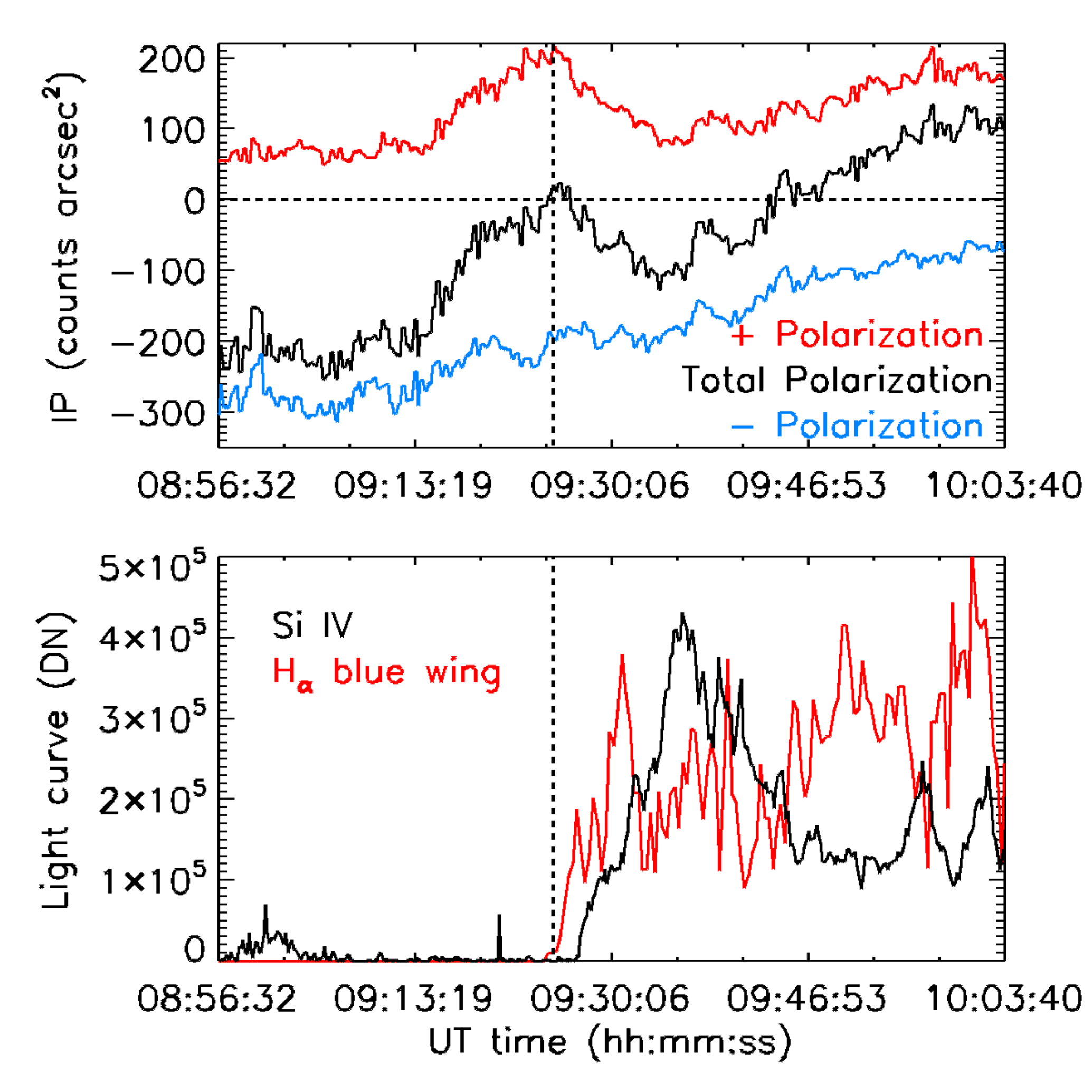}
\caption{Curves obtained from the integration within the solid rectangle plotted in the three panels of Figure \ref{obsfig1}. Top: Integrated polarizaton (IP) where the total polarization is shown in black, negative in blue, and positive polarization in red. The horizontal dashed line indicates the zero value for the polarization. Bottom: Light curve of \SiIV \ (black) and of the \Halpha \ blue wing (red). For both panels, the vertical dashed line marks the instant of the maximum polarization  and the start of the brightenings. \label{obsfig2}}
\end{figure}

%
%
\Needspace{5\baselineskip}
\subsection{Origin and identification of the surge and burst}\label{sec:3.1}

As we said in the introduction, surges are closely related to EFRs. For this reason, we looked for evidences of episodes of magnetic flux emergence and cancellation. Figure~\ref{obsfig2} illustrates the evolution of different integrated 
data within the solid rectangle of the three panels in Figure~\ref{obsfig1} (and accompanying movie). The top panel of Figure~\ref{obsfig2} contains information about the magnetic flux; the black curve corresponds to the signed total flux and shows that, prior to the event, the region is dominated by the negative polarity. Around 09:13 UT, the positive polarity (red curve) increases up to four times its initial value, reaching the maximum value at 09:25 UT (vertical dashed line). This increase is due to positive polarity patches that appear within the rectangle in which we are integrating the flux. After the positive flux reaches its maximum, the total net flux decreases and, around 09:27 UT, first brightenings are observed in the SST \Halpha \ images (see panel C of Figure~\ref{obsfig1}) and also in \CaII. The brightness increase is also evident in the bottom panel of Figure~\ref{obsfig2}, where we plot the light curve of the blue wing of \Halpha\ integrated in the solid rectangle of Figure~\ref{obsfig1}. Co-spatially with the SST brightenings and approximately 2 minutes later, brightenings are also visible in the IRIS SJI 1400: the \SiIV \  burst, which is marked in panel B of Figure~\ref{obsfig1} by the R2 arrow, and whose corresponding integrated light curve is illustrated in the bottom panel of Figure~\ref{obsfig2} in black color. Simultaneously and around the same region where the burst originates, a surge is discernible as a dark and elongated structure consisting of small-scale threads, first in the blue wing and then in the red wing of \Halpha \ and \CaII\ (see R1 arrow in panel C of Figure~\ref{obsfig1}). Those darkenings are visible in the images in wavelength positions corresponding to $\sim$ 20-60 km~s$^{-1}$. This surge has projected lengths ranging from a few Mm up to 8 Mm, while the projected width is around 1 or 2 Mm. After 25 minutes ($\sim$ 09:52 UT), the activity of both the \SiIV \ burst and the surge becomes weak; nonetheless, it is enhanced again almost at the end of the coordinated observation (10:03 UT). This may suggest a recurrent behavior similar to that observed for surges \citep{Schmieder1995, Gaizauskas, Jiang2007, Uddin2012}; and for IRIS bursts \citep{Peter:2014h, Gupta2015}.

\begin{figure}
\epsscale{1.18}
\plotone{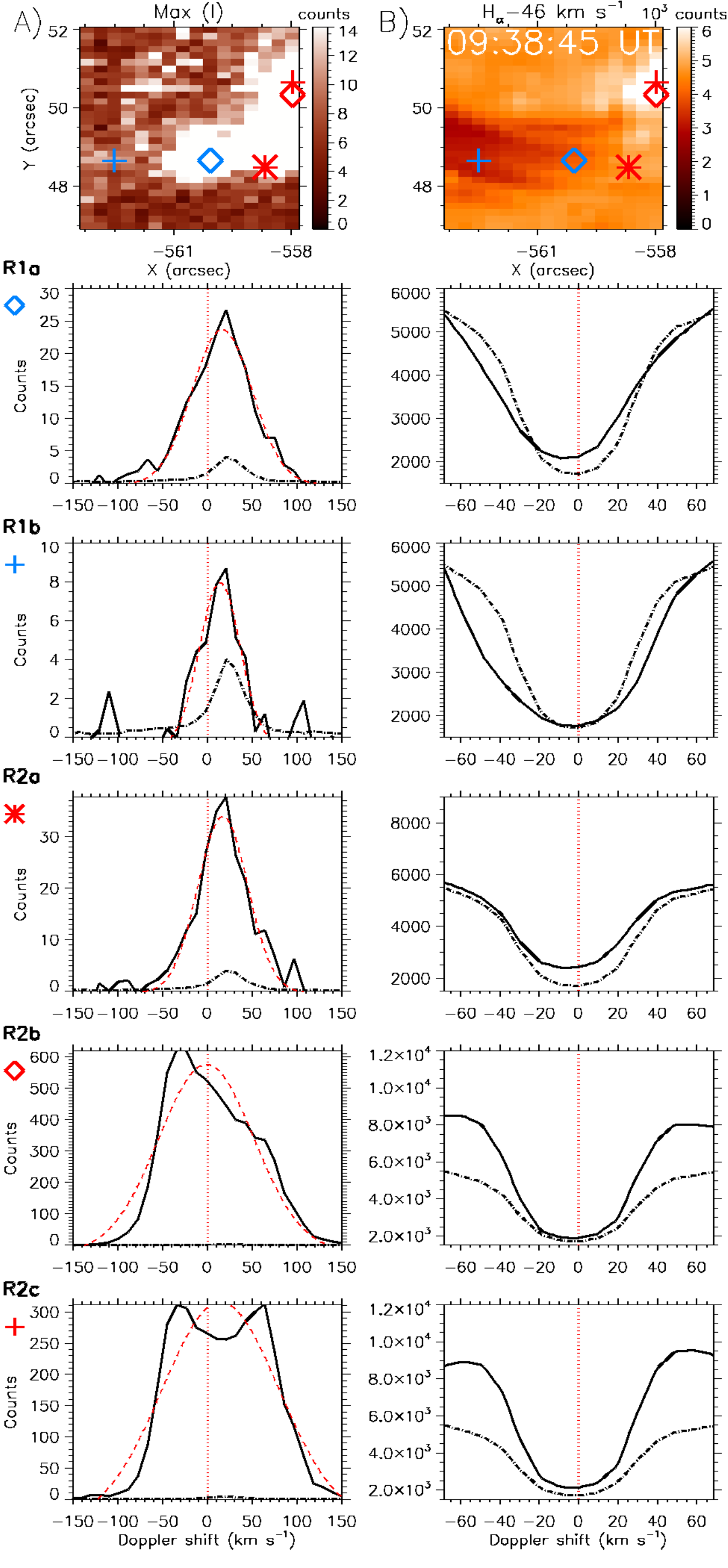}
\caption{(A) Raster intensity peak map of the \SiIV\ 1402.77 $\angstrom$ profile. (B) Image at the blue wing of \Halpha \ (-46~km~s$^{-1}$). Various locations of regions R1 and R2 are indicated by blue and red symbols respectively and their corresponding profiles are shown below panels A and B.  The left column illustrates \SiIV \ 1402.77 $\angstrom$ profiles (black solid line), the Gaussian fit to them (red dashed curve); and the reference average profile (black dash-dotted line). The right column illustrates the \Halpha \  profiles  in the various regions (solid line) and the reference average profile (dash-dotted line). For R1 panels we have averaged four pixels in order to increase the signal to noise ratio of this region. \\
(An animation of this figure is available showing the time evolution of the profiles of each region from $\sim$ 09:30 UT until $\sim$ 09:51 UT.) \label{obsfig3}}
\end{figure}

%
%
\Needspace{5\baselineskip}
\subsection{Spectral properties of the surge and burst}\label{sec:3.2}

In this section we analyze the surge and the accompanying burst, i.e., the structures labelled R1 and R2 in Figure \ref{obsfig1}. To that end, in panels B and C of Figure \ref{obsfig1}, we select an area (dashed rectangle) that coincides horizontally with the region covered by the IRIS raster, namely, $-563\farcs2 \leq x \leq -558\farcs0$ and located vertically between $47\farcs0 \leq y \leq 52\farcs0$. Detailed maps of the intensity peak of \SiIV \ 1402.77 $\angstrom$ and \Halpha\ at -46 km s$^{-1}$ within that rectangle are shown as panels A and B, respectively, in Figure \ref{obsfig3}. Further, the figure contains spectral profiles for the various locations marked with symbols in panels A and B: two for the surge (R1a and R1b, where we have averaged four spatial pixels in order to increase the signal to noise ratio), and three for the burst (R2a, R2b, and R2c). Those profiles are drawn as solid curves and correspond to the lines of  \SiIV \ 1402.77 $\angstrom$ (left column) and \Halpha \ (right column). Furthermore, the left panels contain a Gaussian fit to the \SiIV \ 1402.77 $\angstrom$ profile (red dashed curve), and a reference average profile (black dash-dotted line), which is calculated for the whole FOV and length of the observation.

\begin{figure}
\epsscale{1.20}
\plotone{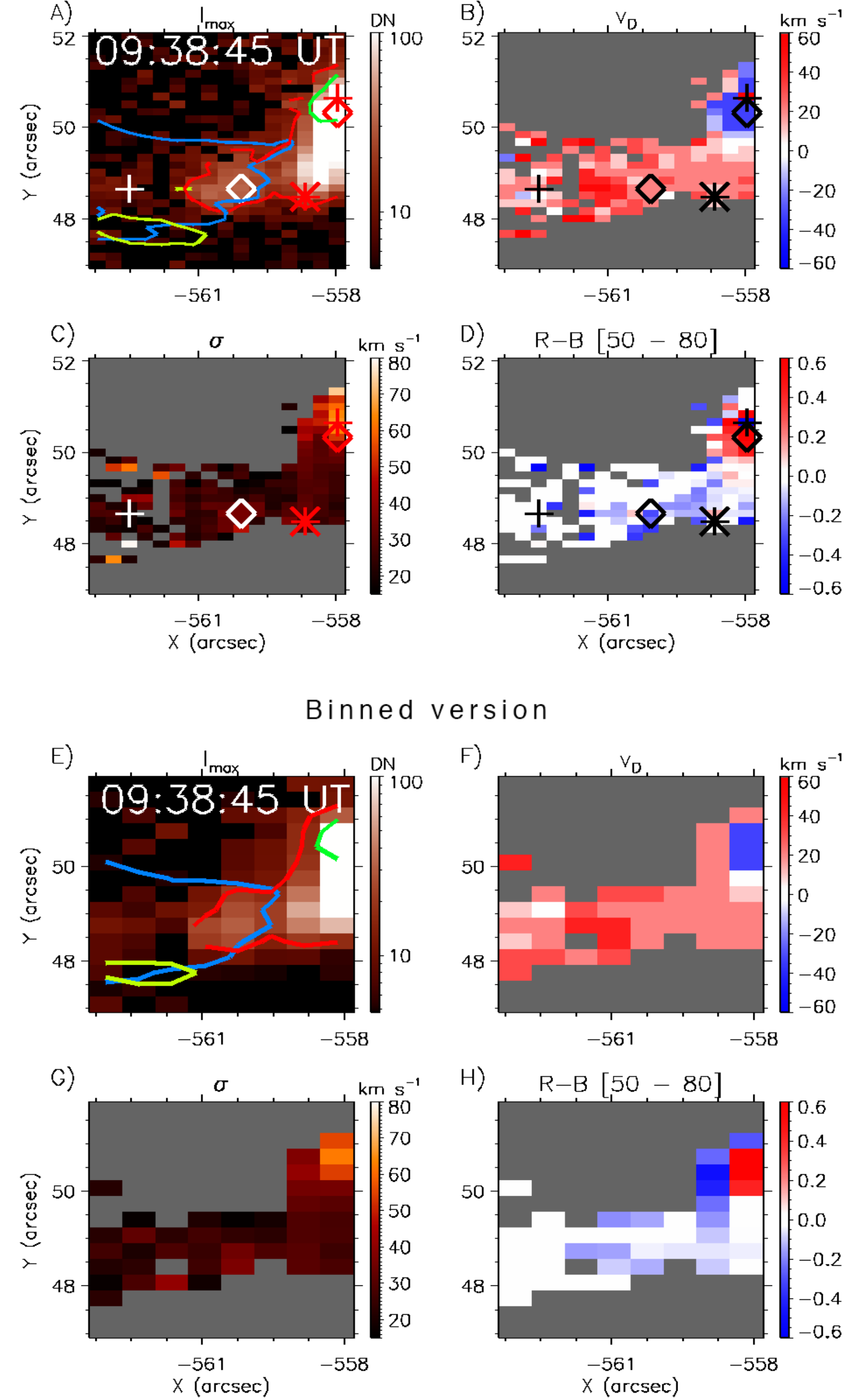}
\caption{The four upper panels are maps of various features of the \SiIV\ spectral profiles. A) Raster intensity peak of the \SiIV\ 1402.77 $\angstrom$ profile at $\sim$ 09:39 UT with overlying contours indicating the burst (red contour where $I_{max} > 16$ DN); the EB (green); the surge in the \Halpha\ blue wing (blue); and in the red wing (yellow). B) Doppler shift $v_D$ in km~s$^{-1}$, where the spectral position corresponding to the maximum intensity is used as the line centroid; C) line width $\sigma$ in km~s$^{-1}$; and D) R-B asymmetry between 50 and 80~km~s$^{-1}$ normalized to the peak intensity. A gray color mask is overplotted in panels B, C and D in places outside the burst and surge, and where the intensity is below 5 counts and $\sigma$ is below 16~km~s$^{-1}$. The four lower panels show the same quantities after a spatial binning by a factor two in each direction. \\
(An animation of this figure is available online containing the time evolution of the panels from $\sim$ 09:30 UT until $\sim$ 09:51 UT.) \label{obsfig4}}
\end{figure}

On the other hand, and to extend the analysis, Figure \ref{obsfig4} contains maps, also at $\sim$ 09:39 UT, of  A) the maximum intensity, $I_{max}$, of the \SiIV \ 1402.77 $\angstrom$ profile; B) the Doppler shift $v_D$; C) the line width $\sigma$; and D) the R(ed)-B(lue) asymmetry, which is the difference between the integral of the two wings of the profile in the same velocity ranges, in this case between 50 and 80~km~s$^{-1}$ \citep[see][for further details about the R-B asymmetry calculation]{De-Pontieu:2009yf,Martinez-Sykora:2011fj,Tian:2011lr}. The variables $v_D$, $\sigma$ and R-B are obtained from the Gaussian fit centered at the spectral position of the peak of the profile. The lower 4 panels contain the same variables but after being binned twice in each direction, thus getting better signal/noise ratio and facilitating the identification of the surge in \SiIV. An animation of this figure is available online.

\Needspace{5\baselineskip}
\subsubsection{R1: the surge}\label{sec:3.2.1}

The region selected for Figure \ref{obsfig3} contains about one third of the surge (R1), which is identifiable as a
dark structure in the \Halpha\ image between $y = 47 \arcsec$ and $y = 50\arcsec$. The \Halpha\ profiles R1a,b underneath show  a dark blue wing with respect to the reference profile. Those two profiles correspond to two different locations of the surge: the brightest point of \SiIV \ 1402.77 $\angstrom$ within the surge (blue diamond, hereafter R1a), and a fixed position around $x=-562.5\arcsec$  and $y=48.5\arcsec$ (blue cross, R1b). In both locations we can identify significant emission in  \SiIV \ that, as far as we know, has not been reported before in the literature. The main properties of this region concerning  \SiIV \ are the following:

\begin{itemize}

\item The emission in \SiIV \ is sporadic and we see it mostly when the surge appears in the blue wing of \Halpha, in particular, between 09:34 UT and 09:41  UT (see animation of Figure \ref{obsfig3}).

\item The \SiIV \  intensity within the surge is weak in comparison to the burst, which could be the reason why it has not been reported before. This is evident in panels A and E of Figure \ref{obsfig4} if we compare the values within the red contour (the burst) with the ones inside the blue contour (the surge). During the event, the ratio between the peak of the \SiIV \ profile and the one of the reference profile (black dash-dotted line in Figure \ref{obsfig3}) is on average around 5.2 in R1a and 1.9 in R1b (see Table \ref{obstable1}), with maximum values of 8.9 and 2.7 respectively.

\item Given its definition, R1a, the brightest \SiIV\ location within the surge, moves with time, 
but we have found that it is mostly located near the region that we have called 
\textit{the footpoint of the surge} in Section \ref{sec:2.1} (see animation of Figure \ref{obsfig3}). It can also be seen that R1a is a small-scale brightening of a few pixels at most. This suggests that IRIS does not completely resolve the spatial structure of R1a. The location of the brightest \SiIV\ emission that is near the footpoints is reproduced in the numerical experiments and could be a characteristic feature of the surges. In fact, in the experiments we also provide a possible reason why R1a is sometimes not so close to the footpoints (Section \ref{sec:4.2.1}).

\item In the observed surge, \SiIV \  shows mainly redshifted profiles (see panels R1a and R1b in Figure \ref{obsfig3} and panels B and F of Figure \ref{obsfig4}). In order to illustrate this quantitatively, we calculate average values of $v_D$ for the two regions R1a and R1b when the \SiIV\ intensity is larger than 5 counts and  $\sigma$ is at least 16~km~s$^{-1}$, i.e., when the \SiIV \ signal is greater than the reference average profile. The results are shown in Table \ref{obstable1}. We see that the \SiIV\ emission within the surge shows redshifts with an average value around 20-23~km~s$^{-1}$. (Note that the relative Doppler velocity is blueshifted around 2~km~s$^{-1}$ only with respect to the reference profile.)

\begin{deluxetable}{cccc}
\tablecolumns{4} \tablecaption{Temporal averages of \SiIV \ 1402.77 $\angstrom$ for: 1) peak intensity $I_{peak}$ with respect to the one of the reference profile $I_{peak,0}$, Doppler shift $v_D$, and line width $\sigma$.
\label{obstable1}} \tablehead{
  \colhead{Region} & 
   \colhead{$< I_{peak}/I_{peak,0} >$} &
  \colhead{$<  v_D  >$ (km~s$^{-1}$)} & 
  \colhead{$< \sigma >$ (km~s$^{-1}$)}} &
  \startdata 
  R1a  &  $5.2 \pm 1.4$   & $20.2 \pm 10.4$  &  $22.9 \pm 5.3$  \\
  R1b  &  $1.9 \pm 0.5$   & $22.7 \pm 20.5$  &  $29.6 \pm 9.4$ 
\enddata
\end{deluxetable}

\item Within the surge, the line width $\sigma$ is around 23-29~km~s$^{-1}$ (see panel C in Figure \ref{obsfig4} and the corresponding average in Table \ref{obstable1}). Those values are approximately a factor 4 greater than the corresponding value of the width just for thermal broadening ($\sigma_{th} = 6.86$~km~s$^{-1}$ taking the temperature formation peak at $\log(T)=4.90$ in statistical equilibrium). We calculate the non-thermal line broadening $\sigma_{nt}$ following \cite{De-Pontieu2015} with 
\begin{eqnarray}
	\sigma_{nt} = \sqrt{ \sigma^2 - \sigma_{th}^2 - \sigma_{inst}^2}, 
\label{eq:sigma_nt}
\end{eqnarray}
where $\sigma_{inst}=3.9$~km~s$^{-1}$. The obtained non-thermal widths, 21.45 and 28.52~km~s$^{-1}$ for R1a and R1b respectively, are on the upper range of the typical values for an active region \citep[see][]{De-Pontieu2015}; nevertheless, we would need more \SiIV \ observations to carry out a statistical analysis in order to know whether this is an intrinsic characteristic of the surges or not.

\item Concerning panel D of Figure \ref{obsfig4}, we find some R-B asymmetry mainly to the blue side in the surge footpoints. This is more evident in panel H of the figure, which contains a version of panel D spatially binned by a factor two in both directions. The found R-B asymmetry imply that a small fraction of the surge plasma is moving with velocities greater than 50~km~s$^{-1}$ relative to the profile peak.

\end{itemize}

\Needspace{5\baselineskip}
\subsubsection{R2: the burst} \label{sec:3.2.2}

The other region we have analyzed is the part of the burst (R2) for which we have IRIS 
spectral profiles. The three lower rows of Figure \ref{obsfig3} show the profiles in three different locations: the brightest \SiIV\ point between $x=-558\farcs67$ and $x=-557\farcs97$, which corresponds to locations within the burst close to the surge footpoints (red asterisk, hereafter R2a); at the location within the burst that has the maximum \SiIV\ emission (red diamond, R2b), and at the brightest point in the \Halpha \ wings at 46~km~s$^{-1}$ within the area of the \SiIV\ burst (red cross, R2c). In the following, we describe the main properties of this region.

\begin{itemize}

\item R2a can be considered  as the transition between the surge and the burst. In this region, the \SiIV \ profiles repeatedly change in time between being blueshifted and redshifted, with small Doppler velocities and line widths of $\pm$30~km~s$^{-1}$ (see the accompanying movies of Figure \ref{obsfig3} and \ref{obsfig4}). In \SiIV, this region is slightly brighter than R1, but weaker than the following ones. The latter can be said also for \Halpha, which is characterized by showing some enhancement, both in the wings and core, but not at the same level than R2b and R2c.

\item Regarding R2b and R2c, the \SiIV \ spectra are characterized by  multicomponent profiles, see the corresponding panels in Figure \ref{obsfig3}; the blue component of the profiles is usually the brightest one, as shown in panel B of Figure \ref{obsfig4} where the peak of the profile is blueshifted; the profiles are broad, see panel C where $\sigma > 60$~km~s$^{-1}$, and they show strong asymmetry to the red, see panel D. The polarization map in Figure \ref{obsfig1} reveals that R2c mostly coincides with the location where the two underlying opposite polarities collide. This situation is often interpreted in the observational literature by assuming that the two components of the \SiIV\ profiles correspond to bi-directional flows due to magnetic reconnection. Further evidence about magnetic reconnection and ejection of plasmoids is provided by \cite{Rouppe2017} who analyse the same IRIS and SST observations but include spectrally resolved \ion{Ca}{2}~K imaging from the new CHROMIS Fabry-P\'erot instrument at the SST.

During the event, the brightest \SiIV\ (R2b) is usually located close to brightest \Halpha\ wing (R2c)
in the upper part (solar Y-axis) 
of the burst in \SiIV , and they can even overlap on a few occasions. Interestingly, between 09:45:30 and 09:46:10, when the surge is visible in the red wing of  \Halpha, R2b moves down in the Y-axis approximately $2\arcsec$ while R2c holds in the same position. This may indicate motion of plasma emitting strongly in \SiIV . The two-component profiles described above and this motion of the brightest point in \SiIV \ within the burst are addressed in Section \ref{sec:4.2.2}.

\end{itemize}

We have also carried out a brief study of the other spectral lines observed by IRIS
and found that R2b and R2c show absorption features superimposed on the \SiIV \ 1393.76~$\angstrom$ 
line that correspond to \ion{Ni}{2} 1393.33 $\angstrom$ redshifted around 13~km~s$^{-1}$. 
We looked for further evidence in the IRIS spectra and also detected absorption of \ion{Ni}{2} 1335.20 $\angstrom$ in the \ion{C}{2} profile. Similar absorption features have been reported in IRIS lines by other authors like \cite{Schmit2014}, \cite{Peter:2014h}, \cite{Kim:2015y}, \cite{Vissers:2015} and \cite{Tian:2016h}, which indicate that there is overlying dense and cold plasma on top of the \SiIV\ source. Concerning the other IRIS lines, there are brightenings in the wings of \ion{Mg}{2} h \& k lines with small enhancement in the cores, which is akin to what was described by \cite{Tian:2016h} about IRIS bursts and point to heating of chromospheric plasma. There is also evidence of emission in the triplet of subordinate lines of \ion{Mg}{2} at 2798.75 and 2798.82 $\angstrom$, indicating an abrupt temperature increase in the lower chromosphere \citep[see][]{Pereira:2015}. The IRIS profiles do not show the forbidden \ion{O}{4} 1401.2 and 1399.8 $\angstrom$ lines, so we cannot use them for density diagnostics as carried out by \cite{Olluri:2013fu}, \cite{Polito2016} or \cite{Martinez-Sykora:2016obs}. There is no evidence, either, of \ion{Fe}{12} at 1349 $\angstrom$, which forms around 1.5 MK. The lack of \ion{O}{4} and \ion{Fe}{12} is likely due to the very short exposure time of this observation (0.5 s) since those lines are fainter and require longer exposure times \citep{De-Pontieu:2014vn}.

With respect to the spectra obtained by SST, in Figure \ref{obsfig3} we see enhanced wings in the \Halpha \ profiles both in R2b and in R2c, which resemble the profiles found by \cite{Kim:2015y} in IRIS bursts. In the associated movie, we can see that the \Halpha\ profiles can show extremely enhanced wings while the core mostly keeps unperturbed as in EBs profiles.

%
\Needspace{5\baselineskip}
\section{Numerical experiment: Results}\label{sec:4}

In this section we analyze synthetic profiles from the numerical experiments introduced in Section \ref{sec:2.2} in order to compare them with the previously described observations, thus attempting to provide theoretical understanding of the phenomena studied. The layout of this section is as follows: 1) a general description of the time evolution of the experiments, and 2) the synthesis and comparison with the IRIS observations.

\begin{figure*}
\epsscale{1.10}
\plotone{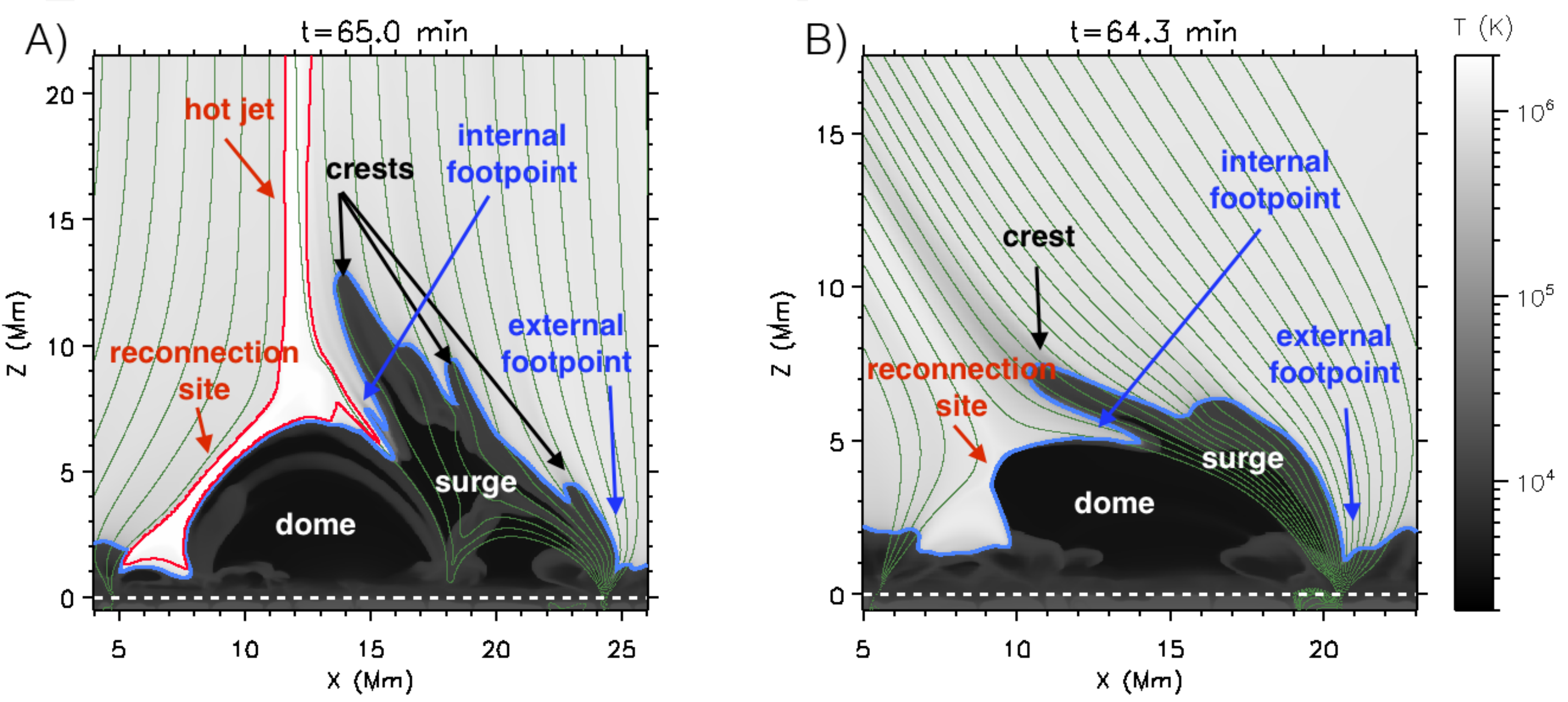}
\caption{2D temperature maps of the surge experiments showing the context and the regions of interest. A) The vertical coronal magnetic field experiment at $t=65.0$ min. B) The slanted coronal magnetic case at $t=64.3$ min. Additionally, magnetic field lines are overimposed in green and temperature contours are added for $T=10^{4.9}$ K (blue) and $T=1.2 \times 10^{6}$ K (red).} \label{simfig1}
\end{figure*}

\begin{figure*}
\epsscale{1.21}
\plotone{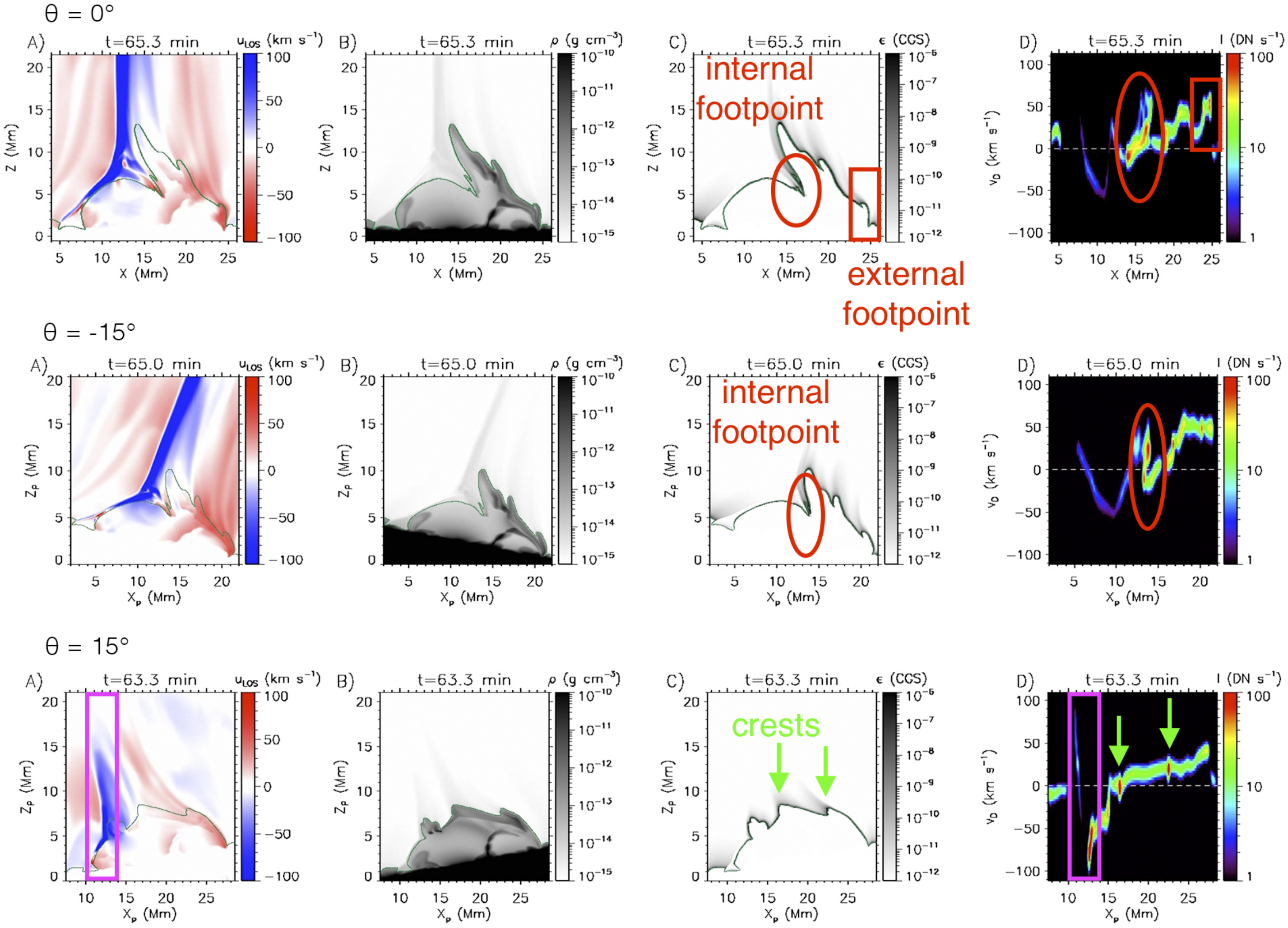}
\caption{2D maps of the vertical experiment at different times for various LOS $\theta$. A) Velocity along the LOS, $u_{_{LOS}}$; B) density, $\rho$; C) emissivity, $\epsilon$, for \SiIV \ in CGS units (erg cm$^{-3}$ sr$^{-1}$ s$^{-1}$); and D), synthetic spectral intensity, $I$ (see Equation \ref{eq:dn}), along the LOS $\theta$.  Temperature of the peak emission for SE ($\log(T) = 4.90$ K) is overplotted as a green contour. In the $\theta \neq 0$ rows, $Z_p$ and $X_P$ are respectively the vertical and horizontal coordinates of the rotated figures.\\
(An animation of this figure is available showing the time evolution of the surge from its origin ($t=61$ min) up to its decay phase ($t=70.7$ min) in the vertical experiment for the three LOS.)} \label{simfig2}
\end{figure*}

\begin{figure*}
\epsscale{1.21}
\plotone{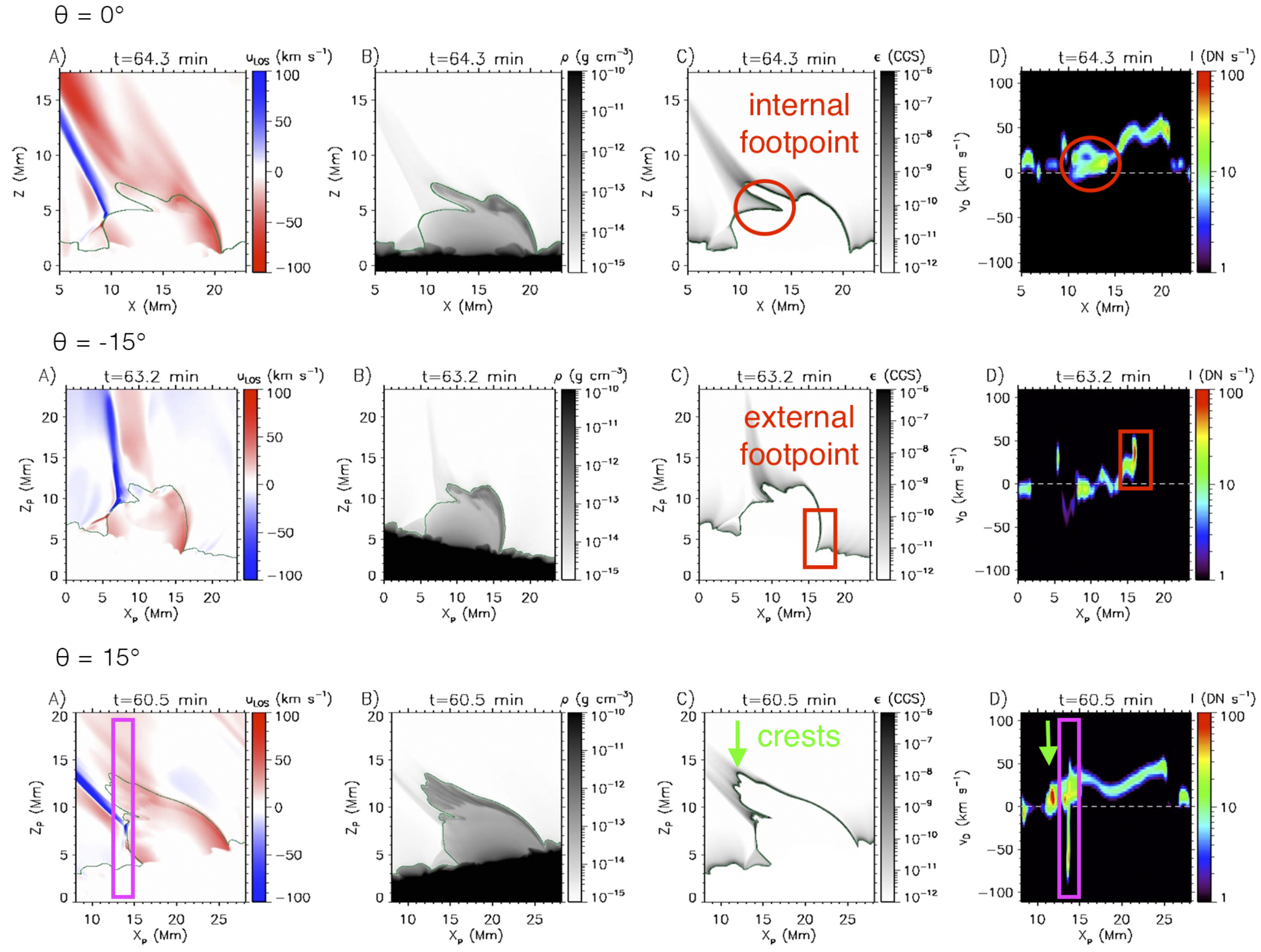}
\caption{Same 2D maps than Figure \ref{simfig2} but for the slanted experiment. \\
(An animation of this figure is also available with the time evolution of the surge from its origin (in this case, $t=58$ min) up to its decay phase ($t=68$ min) for the three LOS.) \label{simfig3}}
\end{figure*}

%
%
\subsection{A brief description of the time evolution}\label{sec:4.1}

The general features of the time evolution of these experiments are qualitatively similar to those described by NS2016. In the present paper we focus on the advanced phases of the evolution, when the magnetized emerged plasma that has expanded into the solar atmosphere collides with the preexisting coronal field, and a surge is obtained as an indirect consequence of magnetic reconnection processes.

Figure \ref{simfig1} shows the context of the numerical simulations through a temperature map for each experiment: panel A, the vertical experiment at $t=65.0$ min, and panel B, the slanted one at $t=64.3$ min. In panel A we can see the emerged plasma with a dome shape between $8 \leq x\leq 18$ Mm; a coronal hot jet, with a classical Eiffel-tower shape (see red contour) and maximum temperatures of $T \sim$ 2-3 MK; and the surge, which is the cool and elongated structure, as in NS2016, located on the right side of the dome. The surge has been produced by a peeling-like mechanism during the non-stationary magnetic reconnection process. This process drags away and ejects plasma from the dome, sometimes as plasmoids, giving rise to structures that resemble crests. The inner part of the surge is composed of elongated thread-like structures of different density that are a consequence of the collision and subsequent merging of plasmoids with the surge. Besides that, there is also a detachment mechanism that splits and separates the surge from the emerged dome (see the details about this process in NS2016). This occurs at a height of around 5 or 6 Mm on the side of the surge closest to the reconnection site, referred to in the following as the \textit{internal side} of the surge. Consequently, we call that region the \textit{internal footpoint} of the surge (as marked by an arrow in the figure). Another interesting region for the comparison with the observations is the base of the external region of the surge: we call this the external footpoint of the numerical surge, and is marked by another arrow in the figure. In panel B, the slanted experiment, the emerged dome is located at $10 \leq x\leq 20$ Mm, and the coronal jet has lower temperatures ($T < 1.2$ MK). In this case the surge does not end up totally detaching from the dome, which could be a consequence of the  magnetic field configuration in the corona. In the follow-up paper \citep{Nobrega-Siverio:2017b}, an in-depth study of the physics taking place in this phenomenon is carried out, including the role of the entropy sources and the consequences of the NEQ ionization of silicon and oxygen for the emissivity.

%
%
\subsection{Analysis of the synthetic spectra}\label{sec:4.2}

With a view to identifying the different observational features explained in the paper (Section \ref{sec:3.2}) with structures and phenomena in the experiments, in the following we analyze the synthetic \SiIV \ 1402.77 $\angstrom$ spectra obtained from the numerical data. The surge in the experiments produces spectral features that resemble those from Region R1 in the IRIS  \SiIV\ observations (Section \ref{sec:3.2.1}). In turn, the region near the magnetic reconnection site, whose location is marked in Figure \ref{simfig1} for both experiments, produces spectral features that resemble the burst (R2) described in Section \ref{sec:3.2.2}. In order to establish the relation between the numerical and observational features for the surge and burst, we use Figures \ref{simfig2} and \ref{simfig3}, which correspond to the vertical and slanted magnetic field experiments respectively.  The different rows correspond to different inclination angles for the LOS. Exploring different inclinations is important since they impact on the obtained intensities as seen in the following. The panels in each row contain A) the velocity along the LOS, $u_{_{LOS}}$; B) the density, $\rho$; C) the \SiIV \ emissivity, $\epsilon$; and D) the synthetic spectral intensity, $I$, derived from Equation (\ref{eq:dn}). In the $\theta \neq 0$ panels, $x_P$ and $z_P$ are, respectively, the horizontal and vertical coordinates of the rotated figures. Note that to facilitate the explanation we are choosing somewhat different times for each row. Animations of the Figures \ref{simfig2} and \ref{simfig3} are provided in the online version.

\subsubsection{The surge}\label{sec:4.2.1}

In the following, we analyze separately various characteristic regions of the surge.

\begin{figure}
\epsscale{1.10}
\plotone{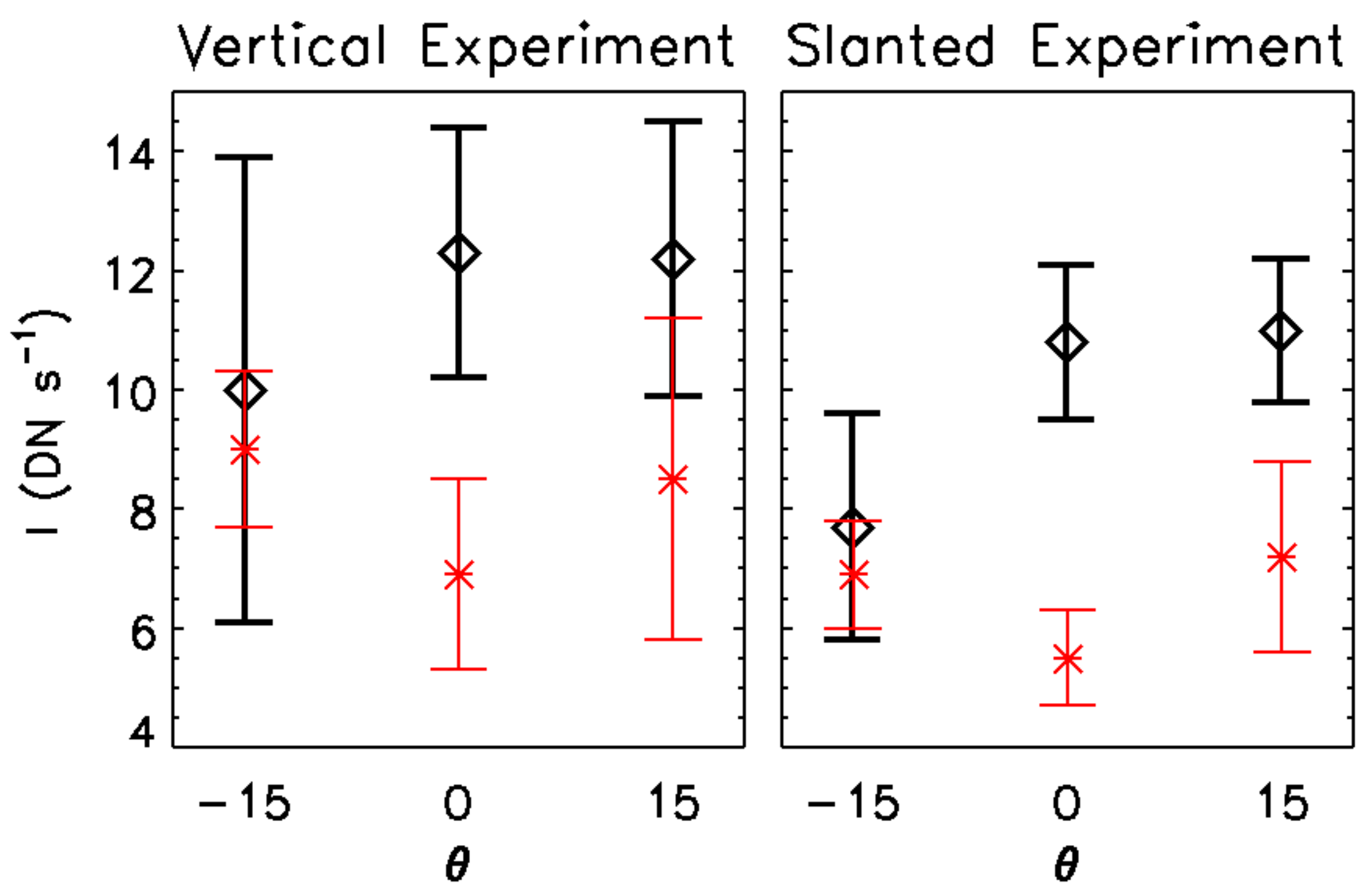}
\caption{Average intensities for the three different LOS of the less bright regions of the surge (black diamonds) and of the \textit{quiet} transition region (red asterisks) for both experiments. The error bars are obtained from the standard deviation.} \label{simfig4}
\end{figure}

\begin{enumerate}[label=(\alph*)]

\item The internal footpoint. For the vertical field experiment (Figure \ref{simfig2}, $\theta = 0 \degree$ and $\theta = -15 \degree$), in panel D at $x \sim 16.5$ Mm and $x_P \sim 13.5$ Mm, respectively, there is a typical spectrum of the internal footpoint obtained through integration along a LOS near and parallel to the transition region at the internal boundaries of the surge. Regarding the slanted experiment (Figure \ref{simfig3}, $\theta = 0 \degree$ row), in panel D, approximately between $11 \leq x \leq 14$ Mm, we can see also a clear example of the spectrum emanating from the internal footpoint of the surge. This region is composed of plasma strongly emitting in \SiIV \, (see panels C) that during the event moves from the reconnection site to the internal footpoint. The plasma here is mostly descending due to the enhanced pressure gradient that pushes it down when going through the post-shock region (see NS2016); as a consequence, the spectra show mainly redshifted profiles. We note that the internal footpoint is usually brighter than the rest of the surge. In the animations of the vertical experiment, we can see that the intensity of this region varies depending on the LOS and can be greatly enhanced when integrating parallel to the transition region of the internal footpoint. 

The features mentioned above fit with the results of the IRIS observation (Section \ref{sec:3.2.1}) for region R1a 
(the brightest region within the surge), which is mostly located around the observational footpoints. The deviations from this behavior are explained in the subsequent items.

\item The external footpoint. Figure \ref{simfig2} shows that in that region (marked in the figure by a red rectangle) we can also find brightenings with intensities between 30 and 112 DN s$^{-1}$. The intensity depends again on the geometry, i.e., it is enhanced when the parallel direction of the transition region of the surge coincides with the LOS. This dependence can be checked by looking at the three different LOS $\theta$ of the same figure. In this region, the Doppler shift $v_D$ has values around 20-60~km~s$^{-1}$ to the red. A similar result is found for the external footpoint of the slanted experiment (see, e.g., the region marked with the circle in the row $\theta = -15 \degree$ of Figure \ref{simfig3}). 

Concerning the observations, the above result strengthens the conclusion of Section \ref{sec:3.2.1} that bright \SiIV\ points within the surge are mostly located at its footpoints. Note that in the current observations we cannot distinguish whether the brightenings of R1a come from the internal or external footpoints.

\item The crests. Representative cases for the spectra of the surge \textit{crests} can be found in the $\theta = 15 \degree$ row of the vertical experiment (marked with green arrows in Figure \ref{simfig2}), which shows: a) a crest in its ascending phase approximately at $x_P=15$ Mm, with intensities around 40~DN~s$^{-1}$ and blueshifted; b) a much brighter one with blue and redshifted components although with small velocities ($x_P=17$ Mm); and c) the last one which is also bright but only shows redshift ($x_P=22.5$ Mm).  At $x_P=13.0$ Mm,  the magnetic reconnection process is leading to another crest as a result of the ejection of plasma upwards with large velocities (see panel A). Regarding the slanted experiment (Figure \ref{simfig3}), the $\theta = 15 \degree$ row shows a crest example between $11 \leq x_P \leq 13$ Mm. In this case, we find that before $t=59.0$ min, the crest is a bright feature on the blue side of the spectrum with absolute $v_D$ of $< 30$~km~s$^{-1}$; nonetheless, as time advances on (e.g, $t=60.5$ min), the whole surge falls down and the crest  then shows redshifted profiles with velocities on the same order. At later stages of the evolution, the crest moves to the right side of the dome and it is not easy to isolate it to integrate the emissivity without going through other interesting sites like the magnetic reconnection region. 

With respect to the comparison with the observations, in Section \ref{sec:3.2.1} we argued that R1a, which was defined
 as the pixel with the brightest \SiIV\ emission within the surge, although mostly located near the footpoints, can sometimes be found further away. Now we realize that those excursions from the observational footpoint region can correspond to instants in which the brightest \SiIV\ emission is produced in the crests of the surge instead of in the footpoints themselves.

\item The rest of the surge. Lets focus now on the less bright regions of the surge. In Figures \ref{simfig2} and \ref{simfig3}, we see that the lowest intensities within the surge are $\sim 10$ DN s$^{-1}$. We would like to explore whether even those regions are intrinsically brighter or not than an average transition region. In order to show this quantitatively, we have carried out a statistical analysis in \textit{quiet} transition regions, i.e., places far away from the locations affected by the magnetic reconnection and the subsequent surge. In those regions, we calculate the average peak intensity during 10 minutes; in the less bright regions of the surge, we average the minimum of the peak intensities within the numerical surge, also during 10 minutes. The resulting intensities and their standard deviations are shown in Figure \ref{simfig4} for the three LOS used in the experiments. The diamonds correspond to the less bright regions of the surge; the asterisks, for the \textit{quiet} transition region. Comparing those values, one could tentatively conclude that even the less bright regions of the surge seems to be slightly brighter than an average transition region; nonetheless, that statement is not fully conclusive given the large size of the standard deviation in some of the cases.

Concerning the IRIS observations, the fixed position R1b (Section \ref{sec:3.2.1}) showed that in \SiIV\ the surge was brighter than the average by a factor of 1.9 (see Table \ref{obstable1}). In fact, in Figure \ref{obsfig4} and associated movie, the whole surge is above the peak intensity of the average profile. In the numerical experiments, the comparison between intensities from a \textit{quiet} transition region and the less bright regions in the surge tentatively appears to be in accordance with the observations, although more statistics from other experiments would be needed for a stronger conclusion.

\end{enumerate}

\subsubsection{The burst}\label{sec:4.2.2}

After analyzing the spectra of the surge, in this section we describe the associated burst. In our numerical experiments we have associated the burst as the reconnection site. This identification is due to the similarity between the observed features and what is explained in the following. Further evidences of the relation between the reconnection site and bursts are found in the recent paper by \cite{Hansteen:2017ib}, where the authors show, through synthesis diagnostics of 3D simulations, that the bursts can be triggered in the mid and low chromosphere owing to the reconnection between different emerging bipolar magnetic fields.

In this paper,  a good example of the burst is found at $x_P=14$ Mm for the slanted experiment (Figure \ref{simfig3}) for $\theta=15 \degree$, which is a LOS parallel to the transition region at the reconnection site. Panel A shows that in that region (pink rectangle) there is a bidirectional flow with large upflow velocities, up to 85~km~s$^{-1}$, and downflow velocities of $< 60$~km~s$^{-1}$. In panel D we notice that the corresponding synthetic profile shares characteristics described in the observation for R2b (the brightest region of the \SiIV\ burst): broad profiles with both red and blue components reaching large Doppler shifts. In the accompanying movie, we see that the blue component is usually brighter than the red one, which is also something found in the observation (see Section \ref{sec:3.2.2}). Turning now to the vertical experiment,  an example of the burst in a direction not parallel to the reconnection site can be found in the $\theta=15 \degree$ row of Figure \ref{simfig2}. Within the pink rectangle in panel A, we see a downflow between $10.0 \leq x_P \leq 11.5$ Mm, and a strong upflow around $11.5 \leq x_P \leq 13.0$ Mm. The synthetic intensity shows large Doppler shifts for both components  but with the blue component being brighter. In both experiments, we see that the plasma emitting in \SiIV\ moves from the reconnection site to the detachment region where the internal footpoint of the surge is formed. This pattern of motion occurs repeatedly, a feature also found in the observation in R2b  (see second bullet point of Section~\ref{sec:3.2.2}). Finally, in the vertical experiment there are also complex profiles due to the plasmoids formation and subsequent cascade. The latter feature is addressed in more detail by \cite{Rouppe2017} using the vertical numerical experiment and comparing it with plasmoids observed with CHROMIS at the SST.

%
\Needspace{5\baselineskip}
\section{Discussion}\label{sec:5}

In this paper, we have used coordinated observations from IRIS and SST to study an \Halpha\ surge and accompanying \SiIV \ burst episode occuring in active region AR12585 on 2016 September 03. During the event, we found for the first time emission of \SiIV \ within the surge and provide details about the properties of the surge and burst. In order to give a theoretical interpretation of the observations, we performed 2.5D RMHD numerical experiments of magnetic flux emergence through the granular cells right below the surface up to the atmosphere. The experiments were carried out with Bifrost using an extra module of the code that calculates the \SiIV \ population in NEQ ionization. During the experiments, surges and bursts are obtained as a direct and indirect consequence of magnetic reconnection processes. We then compute synthetic spectra and compare with the observations.

In the following, we discuss some observational and theoretical features of the surge and burst and also some limitations of the present research.

%
%
\subsection{Observations}\label{sec:5.1}

Analyzing the light curves (Section \ref{sec:3.1}), we have detected that in this case the surge and \SiIV \ burst can be linked spatially and temporally to opposite polarity patches that converge and collide. Moreover, an EB is visible in close vicinity just prior to the start of the burst and remains active during the whole lifetime of the burst. It appears likely that all these different phenomena share a common physical origin.

The most striking feature of the observation is the finding of emission in \SiIV \ within the surge. Previous papers (e.g., \citealp{Kim:2015y} and \citealp{Huang:2017z}) had only addressed the coexistence of surges, seen in \Halpha \ and \CaII \ 8542 $\angstrom$, and \SiIV \  brightenings, but without identifying evidences of \SiIV \ within the surge. Our finding implies that surges, which are traditionally associated with chromospheric lines, have enough impact in the transition region to be detectable. Furthermore, this result together with the one by \cite{Madjarska2009}, where the surge was associated with highly non-Gaussian line profiles in \ion{O}{5} and \ion{N}{5}, qualitatively increases  the interest of studying transition region lines for diagnosis of chromospheric phenomena. The  \SiIV \  emission found in our surge is not strong, its intensity is around a factor 2-5 brighter than a quiet-sun typical \SiIV \ profile, and sporadic: we detect it mostly in the rising phase of the surge, when it is visible in the blue wing of \Halpha \ (Section \ref{sec:3.2.1}). The lack of \SiIV \  in the decay phase of the surge may be due to cooling of the plasma so it is no longer emitting in \SiIV\ or too weak to be detectable in short exposure time observations like the present one. Another interesting feature is that the brightest emission in  \SiIV\  within the domain of the surge is located near its footpoints. Moreover, the profiles are mainly redshifted with asymmetries to the blue. This reflects the complexity of the surge region, where in the same LOS we can find cool surge plasma visible in the blue wing of \Halpha, which indicates rising motion, and hotter plasma emitting in \SiIV, that is mainly descending, while there is a small fraction of large upflow velocities relative to the profile peak. The surge is otherwise of a canonical type: it is visible as an elongated and dark structure first in the blue and then in the red wings of \Halpha \ and \CaII, with projected lengths of several Mm, and Doppler velocities of a few tens of km~s$^{-1}$ (Section \ref{sec:3.1}). Concerning the \SiIV\ burst (Section \ref{sec:3.2.2}), we see typical burst characteristics as reported in the literature: the burst appears in regions on the surface where magnetic flux regions of opposite polarity converge and collide, it has broad profiles due to the presence of blue and red components, absorption features superimposed on the \SiIV \ lines, brightenings in the wings of \ion{Mg}{2} h \& k lines with small enhancement in the cores, and enhanced wings in \Halpha.

A full multi-diagnostic study of surges is challenging since it requires coordinated observations from different instruments. Moreover, full spectral coverage requires the IRIS raster to coincide both spatially and temporally with the region where the events are occurring, which reduces the number of available observations for a statistical analysis. In our case, a severe limitation is the short exposure time of this observation (0.5 s), so lines  
like \ion{O}{4}~1401.2 and 1399.8~\AA\ cannot be used for density diagnostics since they require longer exposure times to be detectable \citep{De-Pontieu:2014vn}. Furthermore, the raster does not cover the whole surge, 
so we do not have a complete data set of the surge for a more complete comparison and questions about the emissivity in \SiIV \ at the upper part of the surge remain open.

%
%
\subsection{Numerical experiments}\label{sec:5.2}

In the paper, \SiIV\ spectral synthesis has been used for the first time to study surges (Section \ref{sec:4.2.1}) and the associated burst (Section \ref{sec:4.2.2}). We have found that our models are able to reproduce some of the main features of the observations as summarized in the following: 

\begin{itemize}

\item The brightest points in \SiIV\ within the surge correspond to the location around the footpoints of the surge. The excursions of the brightest patch from the observed surge footpoints are identified with the crests in the surge.

\item The synthetic \SiIV\ profiles obtained in the experiment for the surge are mainly redshifted and the values fit with the observed Doppler shifts.

\item The less bright regions of the surge in \SiIV\  seem to be brighter than an average transition region and match tentatively with what was found in the observations.

\item Our experiments provide a temporal and spatial relation between the surge and the burst. Both are a natural consequence of magnetic reconnection between emerged plasma and the preexisting coronal magnetic field.

\item The motions observed with IRIS from the brightest region in the \SiIV\ burst to the surge footpoints can be identified in the numerical experiments through plasma strongly emitting in \SiIV\ that moves from the reconnection site to the internal footpoint of the surge.

\item The two-component profiles of the observed burst are characterized by large velocities and by the fact that the blue component is usually brighter than the red one: this is also found in the numerical experiments. The main difference between theory and observations concerns the burst intensity, which is substantially larger in the observations. A possible reason for the discrepancy is the lack of numerical resolution \citep{Innes:2015,GuoL:2017}.

\end{itemize}

In the numerical experiments we see that the orientation of the LOS  plays an important role for the diagnosis. The resulting spectral intensity in the transition region lines can be enhanced by an important factor when the LOS runs parallel (or almost) to the local transition region. This is something to take into account for future interpretations from observations.

In spite of the good agreement with the observations, our current numerical experiments are not free from limitations, which are related to the challenging conditions in chromospheric plasma and associated radiative transfer. The experiments lack non-equilibrium ionization of hydrogen and helium, which could impact on the properties of the emerged plasma and subsequent surge (\citealt{Leenaarts:2007sf,golding2014}). They also lack ambipolar diffusion, which has recently been shown to be key to obtaining related phenomena like type II spicules \citep{Martinez-Sykora:2017sc}. Moreover, our experiments are 2.5D, but 3D modeling would be necessary to capture the full complexity of those phenomena, as shown recently for bursts and EBs by \cite{Hansteen:2017ib}, and to be able to accurately synthesize the \Halpha\ line \citep{Leenaarts:2012cr}. Proper \Halpha\ synthesis is important, e.g., to identify the numerical threads in the inner part of the surge (Section \ref{sec:4.1}) with observed thread-like structures \citep{nelson2013, Li2016}, or to provide a theoretical explanation for the shocks associated with surges \citep{YangH:2014}.

%
\ \vspace{-2mm}

\acknowledgments We gratefully acknowledge financial support by the Spanish Ministry of Economy and Competitiveness (MINECO) through projects AYA2011-24808 and AYA2014-55078-P, as well as by NASA contract NNG09FA40C (IRIS) and NASA grants NNH15ZDA001N-HSR and NNX16AG90G. IRIS is a NASA small explorer mission developed and operated by LMSAL with mission operations executed at NASA Ames Research center and major contributions to downlink communications funded by ESA and the Norwegian Space Centre. The Swedish 1-m Solar Telescope is operated on the island of La Palma by the Institute for Solar Physics (ISP) of Stockholm University in the Spanish Observatorio del Roque de los Muchachos of the Instituto de Astrof\'isica de Canarias. We also acknowledge the computer resources and assistance provided at the MareNostrum (BSC/CNS/RES, Spain) and TeideHPC (ITER, Spain) supercomputers. Finally, the authors are grateful to Gregal Vissers for his constructive comments during the Hinode-11/IRIS-8 science meeting.

%
\bibliographystyle{apj} \bibliography{collectionbib}

\end{document}